\documentclass[prc,showpacs,floatfix,preprint]{revtex4}

\usepackage{amsmath,amsfonts,amssymb,amsthm,amsxtra,stmaryrd}
\usepackage{graphicx}% Include figure files
\usepackage{dcolumn}% Align table columns on decimal point
\usepackage{bm}% bold math

\usepackage{acronym}  % expand acronyms once

\begin{document}

\acrodef{ALACM}		{Adiabatic Large Amplitude Collective Motion}
\acrodef{CHFB}		{Constrained Hartree-Fock-Bogoliubov}
\acrodef{HB}		{Hartree-Bogoliubov}
\acrodef{HF}		{Hartree-Fock}
\acrodef{HFB}		{Hartree-Fock-Bogoliubov}
\acrodef{LHA}		{Local Harmonic Approximation}
\acrodef{LHE}		{Local Harmonic Equation}
\acrodef{QRPA}		{Quasiparticle Random Phase Approximation}
\acrodef{PPQ}		{Pairing Plus Quadrupole}

%\renewcommand{\mathbb}{\varmathbb}
% dirac notation
\newcommand{\norm}[1]{\left| #1 \right|}
\newcommand{\bra}[1]{\left\langle #1 \right|}
\newcommand{\redbra}[1]{\left\langle #1 \right\|}
\newcommand{\ket}[1]{\left| #1 \right\rangle}
\newcommand{\redket}[1]{\left\| #1 \right\rangle}
\newcommand{\braket}[2]{\left\langle #1 | #2 \right\rangle}
\newcommand{\bracket}[3]{\left\langle #1 #2 #3 \right\rangle} % dirty hack --- forget it!!!
\newcommand{\dotprod}[2]{\: #1 \, \cdot \, #2 \:}
\newcommand{\expectation}[1]{\left\langle #1 \right\rangle} 
\newcommand{\reducedbracket}[3]{\left\langle #1 \left\| #2 \right\| #3
\right\rangle} 
\newcommand{\trace}[1]{{\text{Tr}}\left[ #1 \right]} % trace
\renewcommand{\exp}[1]{e^{ #1}} % 
\renewcommand{\det}[1]{\text{det}{\left[ #1 \right]}} % based on ams

\preprint{}

\title{Removal of Spurious Admixture in a Self-consistent Theory of
Adiabatic Large Amplitude Collective Motion}

\author{Toby D.~Young} \email{young@theory.phy.umist.ac.uk}
 \author{Niels R.~Walet} \email{niels.walet@manchester.ac.uk}
 %\altaffiliation[Also at ]{Physics Department, XYZ University.}
 \affiliation{School of Physics and Astronomy, The University of Manchester, Manchester, M60 1QD, U.K.}


\date{\today}% It is always \today, today,

\begin{abstract}

In this article we analyse, for a simple model, the properties of a
practical implementation of a fully self-consistent theory of
adiabatic large-amplitude collective motion using the local harmonic
approach.  We show how we can deal with contaminations arising from
spurious modes, caused by standard simplifying approximations. This is
done both at zero and finite angular momentum.  We analyse in detail
the nature of the collective coordinate in regions where they cross
spurious modes and mixing is largest.

\end{abstract}

\maketitle

\section{Introduction}
One of the long-standing concerns in nuclear structure theory is to
understand how the collective properties of a microscopic many-body
system emerge from the behaviour of its many quantal constituents
\cite{book:BlaizotRipka1986,book:RingSchuck1980}.  
Bohr;s liquid drop model can be used to describe
many collective modes  \cite{article:BohrMottleson1955}
and still provides us with the concepts used in most
discussions of collective motion. Clearly, a more microscopic method
to describe such modes has been sought and many methods exist
that try to describe collective motion.

One method commonly used is based on a mean-field approach, the
\ac{CHFB} method \cite{book:RingSchuck1980}.  Here a collective energy
surface is generated by mapping the energy expectation value as a
function of the expectation values of a small number of one-body
operators (mean-field constraints).  The principal weakness of
\ac{CHFB} is that the collective motion of the nucleus is determined
without any dynamical criterion for choosing one particular constraint
(collective operator) over another. Often the choice is inspired by the
Bohr Hamiltonian, and one uses only quadrupole and pairing
operators. In nuclear physics the first applications of the \ac{CHFB}
were to the dominant surface vibrations
\cite{article:BarangerKumar1965,article:BarangerKumar1968,article:BarangerKumar1968b,
article:BarangerKumar1968c,article:BarangerKumar1968d} which called
attention to the need for a more meaningful theory.

Ideally the method chosen to tackle this problem should allow the
dynamics of the system to evolve through the microscopic Hamiltonian
without the intervention of {\it ad hoc} elements forcing the system
into a specific mode of oscillation. One possibility, followed in this
article and discussed in detail in the review article
\cite{review:DangKleinWalet2000}, is a method that determines
collective motion by following the normal modes of oscillation. This
incorporates methods such as the \ac{QRPA} \cite{book:RingSchuck1980}.
Thus the low-lying physical excitations of the system using the
small-amplitude harmonic limit (i.e., \ac{QRPA}) are used to
self-consistently determine the constraining operator. Restricting to
the case of a single collective excitation, we then describe a
collective path through the energy surface generated by the
Hamiltonian.

Initially \cite{review:DangKleinWalet2000} most applications ignored
pairing, and they used versions of the \ac{HF} method. In recent
years, three different models including pairing, and thus requiring
the \ac{HFB} mean-field approximation, have been investigated. The
first \cite{article:NakatsukasaWalet1998c,article:NakatskasaWalet1998}
describes a two crossing levels, with an additional pairing
interaction.  One problem first encountered in that work was that of
dealing correctly with spurious modes arising from spontaneously
broken symmetries within the mean-field approximation even though for
this model one can solve this problem exactly, unlike the problems
discussed below. The second
\cite{article:NakatskasaWalet1998b,article:KobayasiNakatsukasaMatsuoMatsuyanagi2000}
is a microscopic $O(4)$ model with pairing and quadrupole interactions
which focuses on shape coexistence phenomena in a schematic way.  The
third set of applications
\cite{article:AlmehedWalet2003,article:AlmehedWalet2004}
uses the \ac{PPQ} model for semi-realistic nuclei using techniques
developed earlier \cite{article:NakatsukasaWaletDang1999}. These
calculations typically use model spaces consisting of two major shells
for both neutrons and protons, leading to large dimensions for the
\ac{QRPA}, and extremely time-consuming calculations. A balance is
found by introducing a basis of operators which gives a truncated
expression of the \ac{QRPA} in such a way that it gives a reasonably
accurate approximation of the low-lying excitations
\cite{article:AlmehedWalet2002,article:KNMM2005}. This method has recently been
extended to finite rotational frequencies
\cite{article:AlmehedWalet2004}.

In this paper we do not make such a truncation, but we will
investigate the nature of the approximations made before doing the
projection on a small basis of operators. To this end we employ a
single $j$-shell \ac{PPQ} model.  The main difficulty encountered here
is associated with admixtures of spurious modes (Nambu-Goldstone
modes) arising in the formalism as vacuum-degenerate solutions in the
excitation spectra which do not correspond to physical
excitations. Though at extrema on the energy surface spurious modes do
not mix with other modes, this is not necessarily true at other points
on the energy surface, especially after additional approximations have
been made. Since spurious modes do not correspond to physical
excitations, any mixing of these modes with other excitations --- we
will refer to this phenomenon as `spurious admixture' --- can
potentially lead the system away from the collective path.  In order
to ensure that spurious modes behave correctly far from stable
equilibrium we need to modify our algorithm slightly, and the study of
these modifications is the main focus of this article.

This article is organised as follows: First, in section \ref{sec:lha}
we give a brief overview of the \ac{LHA} followed in this article.  We
also give a description of our projection technique to remove spurious
admixture from the formalism.  Secondly, in section
\ref{sec:single_j_ppq_model}, following the lead given by references
\cite{article:AlmehedWalet2003,article:AlmehedWalet2002,article:NakatsukasaWalet1998c,article:NakatskasaWalet1998b,article:KobayasiNakatsukasaMatsuoMatsuyanagi2003,article:AlmehedWalet2004,article:NakatskasaWalet1998,article:KobayasiNakatsukasaMatsuoMatsuyanagi2000,article:KNMM2005,letter:NakatsukasaWaletDang1999,article:KleinWaletDang1994,article:NakatsukasaWaletDang1999},
the single $j$-shell \ac{PPQ} model is presented and solved.  We
present our main results from testing our methods of removal of
spurious admixture, as well as discussing a complementary
investigation of the effect of the pairing strength on the collective
potential.  In the next section, Sec.~\ref{sec:single_j_ppq_model-frv}
we then study the behaviour of our techniques at finite angular
momentum.  Finally, a summary and outlook is given in section
\ref{sec:summary_and_outlook_single_jshell}.

\section{Local harmonic approximation}
\label{sec:lha}

In this section we give a brief overview of the \ac{LHA} as used in this
article; for a more detailed exposition of the theory see
Ref. \cite{review:DangKleinWalet2000}. We also give a description of
two methods to remove spurious admixtures from the formalism.

The \ac{LHA} starts from assuming a non-relativistic many-body
Hamiltonian $H_{\text{nuclear}}$ that is capable of describing the
low-energy properties of nuclear structure.  By means of the
time-dependent mean field approximation, the quantum problem is turned
into a classical Hamiltonian problem and the description of collective
motion can now be formulated fully as a classical decoupling problem.
We now try to find collective coordinates such that the mixing between
collective and non-collective degrees of freedom is small.  
The Hamiltonian contains 
$N$ --- the number of quasi-particle states squared ---
canonical pairs of coordinates 
$\xi^\alpha\in\xi=\{\xi^1,\hdots,\xi^N\}$ and momenta
$\pi_\alpha\in\pi=\{\pi_1,\hdots,\pi_N\}$ in the Hamiltonian
$\mathcal{H}\equiv\mathcal{H}(\xi,\pi)$.  In the adiabatic limit,
valid when the collective motion is slow, we
Taylor expand the Hamiltonian up to second order in momenta. We find
\begin{eqnarray}
\mathcal{H}(\xi,\pi)= V(\xi)+\frac{1}{2}\pi_\alpha
B^{\alpha\beta}(\xi)\pi_\beta \quad.
\end{eqnarray}
%where
%\begin{eqnarray}
%B^{\alpha\beta}=\left.\frac{\partial^2\mathcal{H}}
%{\partial\pi_\alpha\partial\pi_\beta}
%\right|_{\pi=0}\quad.
%\label{eqn:classical_ham}
%\end{eqnarray}
The dynamics of the system are thus characterised by the point
function $V(\xi)$ and by the reciprocal mass tensor $B^{\alpha\beta}$,
which also plays the role of a metric tensor.

We now change representation from the initial set of coordinates and
momenta ($\xi,\pi$) to a new set $(q,p)$ through point transformation,
{\it i.e.}, $q=f(\xi)$,
in the hope that we can approximately decouple the collective from
non-collective motion.  Using the standard Einstein convention ---
where a comma denotes a partial derivative $f_{,
\alpha}\equiv\frac{\partial f}{\partial \xi^\alpha}$ --- the point
transformation has the form
\begin{eqnarray}
\xi^\alpha= g^\alpha(q^1,\hdots,q^N)\equiv g^\alpha(q)\quad,&\quad
\pi_\alpha= f^\mu_{,\alpha}p_\mu\quad,
\end{eqnarray}
where $g$ is the inverse of $f$.
The Hamiltonian takes the new form
\begin{eqnarray}
\mathcal{H}(\xi,\pi)=\mathcal{\bar{H}}(q,p)=
\bar{V}(q)+\frac{1}{2}p_\mu\bar{B}^{\mu\nu}(q)p_\nu\quad.
\label{eqn:classical_ham2}
\end{eqnarray}
Here a bar denotes the transformed potential ($\bar{V}=V(\xi(q))$)  and the
transformed mass tensor ($\bar{B}$).

The basic equation of the problem can now be derived by looking at the
dynamical fluctuations at an arbitrary point in coordinate space,
(the \ac{LHE})
\begin{eqnarray}
M^{\alpha}_{\beta}f^\mu_{,\alpha}&=&\Omega^2_\mu f^\mu_{,\beta}\quad,
\label{eqn:local_qrpa}\\
M^{\gamma}_{\beta}&=&\bar{V}_{;\alpha\beta}B^{\beta\gamma}\quad,
\label{eqn:qrpa_mat}
\end{eqnarray}
where a semi-colon denotes a covariant derivative.  The second
equation states that the force should be in the direction of one of
the eigenvectors,
\begin{equation}
V_{,\alpha}=\lambda f^1_{,\alpha}.
\end{equation}
For exact decoupling this direction is (co-)tangential to the
collective path, but we construct an algorithm for finding cases for
approximate decoupling by not imposing this as a condition. Instead,
as discussed in Ref.~\cite{review:DangKleinWalet2000}, we do have a
measure for the quality of collective path based on the deviations
from this criterion.

In the equations above, the covariant derivative is defined using the metric 
$B^{\alpha\beta}$ provided by the kinetic energy.
%as
%\begin{eqnarray}
%V_{;\alpha\beta}&\equiv& V_{,\alpha\beta}-\Gamma^\gamma_{\alpha\beta}V_{,\gamma}\quad,
%\end{eqnarray}
%where the Christoffel symbol $\Gamma^\gamma_{\alpha\beta}$ is defined
%in the usual way by
%\begin{eqnarray}
%\Gamma^\gamma_{\alpha\beta}&=&B^{\gamma\delta}
%(B_{\delta\beta,\gamma}
%+B_{\delta\gamma,\beta}
%-B_{\beta\gamma,\delta})\quad.
%\end{eqnarray} 
The calculation of the covariant derivative requires the inversion of
the mass matrix, which is fraught with numerical difficulties when $B$
has zero eigenvalues. Therefore we shall make the usual assumption
that we can ignore the effects of these curvature terms, i.e,
$\bar{V}_{;\alpha\beta}\sim\bar{V}_{,\alpha\beta}$.

\subsection{Method of removing spurious admixture}
\label{sec:removal_of_spurious_admixture}

It can be shown that for the \ac{HFB} theory considered here, the
\ac{LHE} is a simple generalisation of the QRPA to non-equilibrium
states. Thus, as in the usual equilibrium \ac{HFB} plus
\ac{QRPA} framework spurious modes arise as artifacts of
mean-field symmetry breaking.  In a numerically exact calculation,
without approximations, such modes decouple from the problem, and we
so not have to consider them in detail. When approximations are made,
especially the simplification arising when neglecting the covariant
terms in the derivatives, the situation is different and we get spurious
admixtures.  Since these admixtures do not correspond to physical
excitations we must remove their components completely from the
formalism without altering the meaning of the collective coordinate.
In all cases we know the spurious operator; it may correspond to a
coordinate or momentum, but we do not normally know the conjugate
variable.

Our method is easily generalised to include any number of spurious
coordinates and is therefore not specific to the model we solve in
this article. As we will
see, the main difficulty with this approach is that we do not know
the conjugate variables and we will
therefore have to devise a way to approximate an operator for the
momenta in order to complete our projection method (this is discussed
below).

The first step is thus the construction of a set of approximate
conjugate variables. This is best illustrated for the particle number,
although other operators are dealt with similarly.

Thus we have $q_{,\alpha}=\mathcal{N}_{,\alpha}$, and its conjugate
momentum $p^\alpha=\phi^\alpha$.  We are looking for operators that
satisfy the conditions %(\ref{eqn:nonzero_eigenvalue_N}).
\begin{eqnarray}
B^{\alpha\beta}\mathcal{N}_{,\beta}=0\quad,\quad
\mathcal{\phi}^{\alpha}V_{,\alpha\beta}\propto\mathcal{N}_{,\alpha}\quad.
\label{eqn:nonzero_eigenvalue_N}
\end{eqnarray}

Let us now turn to the canonical coordinates and momenta and the
question of how to define the momentum variable $\phi$ conjugate to
the particle number $\mathcal{N}$. Naive inversion of the second equation
in  (\ref{eqn:nonzero_eigenvalue_N}) gives
\begin{eqnarray}
\phi^\beta\propto\mathcal{N}_{,\alpha}(V_{,\alpha\beta})^{-1}\quad.
\label{eqn:phi}
\end{eqnarray}
We thus approximate $\phi$ by 
\begin{eqnarray}
\phi^\alpha=\sum_j\frac{1}{\epsilon_j}e^\alpha_j
\mathcal{N}_{,\alpha}e^j_\alpha\quad,
\end{eqnarray}
where $e^j_\alpha$ are components of the eigenvectors and $\epsilon_j$
the eigenvalues of the potential matrix.  Unfortunately
$V_{,\alpha\beta}$ is usually singular due to zero eigenmodes of the
\ac{QRPA} matrix, and we need to replace (\ref{eqn:phi}) by a singular
value decomposition.  To this end we remove those terms in the sum
corresponding to eigenvalues that are close to zero $\epsilon_j\sim0$.
In practise, the size of the excluded zone is found through trial and
error.

Finally, we need to ensure that the operators arising from our construction
are all conjugate to each other:
 In
a generalised representation we have a set of canonical conjugate
coordinates and momenta connected with spurious modes
\begin{subequations}
\label{eqn:coord_momenta-frv}
\begin{eqnarray}
&q^i_{,\alpha}&\quad,\quad\forall\,i\in\{1,2,\hdots,N_S \}\quad,\\
&p_j^\alpha&\quad,\quad\forall\,j\in\{1,2,\hdots,N_S \}\quad,
\end{eqnarray}
\end{subequations}
where $N_S$ denotes the total number of spurious modes. For well defined
coordinates and momenta we have  $q^i_{,\alpha}p_j^\alpha=\delta^i_j$, but
our construction above usually does not satisfy this (partially due to neglect
of covariant terms, and partially due to numerical problems).

We start by defining the overlap matrix
\begin{eqnarray}
O^i_j=q^i_{,\alpha}p^\alpha_j\quad,
\end{eqnarray}
which, had we have done the computation consistently, would be the
identity matrix corresponding to the Poisson brackets between
canonical coordinates and momenta. To achieve canonicity we now
diagonalise the overlap matrix
\begin{eqnarray}
(S^{-1})^k_lO^i_jS^j_l=\delta^k_l\tau_l\quad,
\end{eqnarray} 
where $S$ are a set of orthonormal eigenvectors and $\tau_l$ are the
eigenvalues of the system. This allows us to define a new `tilde'
basis
\begin{eqnarray}
\tilde{q}^i_{,\alpha}&=&(S^{-1})^i_jq_{,\alpha}^j\quad,\\
\tilde{p}_i^\alpha&=&\frac{1}{\tau_j}S^i_jp^\alpha_i\quad,
\end{eqnarray}
satisfying the orthonormal condition
$\tilde{q}^i_{,\alpha}\tilde{p}_j^\alpha=\delta^i_j$, used in the
definition of the projection operator
\begin{eqnarray}
P^\beta_\alpha=\delta^\beta_\alpha-\tilde{q}^i_{,\alpha}\tilde{p}^\beta_j\quad,
\end{eqnarray}
The projection matrix is now applied to the potential and mass matrices,
\begin{eqnarray}
\tilde{V}_{,\alpha\beta} =P^{\alpha^\prime}_\alpha
V_{,\alpha^\prime\beta^\prime}P_\beta^{\beta^\prime}\quad,\quad
\tilde{B}^{\alpha\beta}=
P^\alpha_{,\alpha}B^{\alpha^\prime\beta^\prime}P^\beta_{\beta^\prime}\quad.
\end{eqnarray}
The related QRPA eigenvalue problem has now a number of zero eigenvalues,
and the spurious modes do not mix with with any other ones.

\section{Model and results}
\label{sec:single_j_ppq_model}

In this section we solve the single $j$-shell \ac{PPQ} model for
\ac{ALACM} using the methods discussed above to remove spurious
admixture. The \ac{PPQ} model we use here is a good starting point
despite its limitations as it is relatively simple compared to models
with more realistic forces and allows us to choose a conveniently
small basis set with which to work. Furthermore, we would like to
`bridge the gap' between simpler applications starting from a partial
Hamiltonian and semi-realistic descriptions of real atomic nuclei (see
discussion above).

Following reference
\cite{article:MangSamadiRing1976} the \ac{PPQ} Hamiltonian will be
treated in the \ac{HB} framework.  We start with the traditional
single $j$-shell \ac{PPQ} Hamiltonian
\begin{eqnarray}
\hat{H} =
-\frac{1}{2}\kappa\sum_M(-1)^M:\hat{Q}_{M}\hat{Q}_{-M}:
-\chi\hat{P}^\dagger\hat{P}\quad,
\label{eqn:schematic_op_ham}
\end{eqnarray}
which is a straightforward two-body interaction Hamiltonian.  Here
colons denote normal ordering and $\kappa$ and $\chi$ are the coupling
strength of the quadrupole and pairing interactions, respectively.  We
can only impose correct particle number on average, which is done by a
Lagrange multiplier.  When looking at non-zero angular momentum we 
impose an additional constraint for the expectation value of $J_x$.

The quadrupole and pairing operators are to be written in the simplest
form possible in the original particle basis
\begin{eqnarray}
\hat{Q}_{M} &\propto&  \sum_{mm^\prime}(-)^{j-m-M}
\left[\begin{array}{c@{\:}c@{\:}c@{\:}} j & j & 2 \\ m & -m^\prime & M\\\end{array}\right]
c^\dagger_m c_{m^\prime}\quad,
\label{eqn:qpole_operator}\\
\hat{P} &=& \sum_{m>0} (-)^{j-m} c_{-m}c_m \quad,
\label{eqn:pairing_destruction_operator}
\end{eqnarray}
where $c^\dagger_m$ and $c_m$ are creation and destruction operators
on the single-particle state $\ket{m}$, and the object in square
brackets denotes a Clebsch-Gordan coefficient.  Since we are working
in a single shell, we can absorb the reduced matrix element of the
quadrupole operator into $\kappa$, and use an equality in
Eq.~(\ref{eqn:qpole_operator}) We use the definitions
\begin{equation}
q_M=\bracket{}{\hat{Q}_M}{}\quad,\quad q_{\pm 1}=0,\quad
p_0=\bracket{}{\hat{P}^\dagger}{}=
\bracket{}{\hat{P}}{}\quad,
\end{equation}
where angular brackets denotes the expectation value.  We will use the
Hill-Wheeler coordinates \cite{article:HillWheeler1953}
\begin{eqnarray}
\beta=\pm{\sqrt{q_0^2+q_2^2}}\quad,
\gamma=\arctan\left(\sqrt{2}\frac{q_2}{q_0}\right)\quad.
\end{eqnarray}

We use the standard \ac{HFB} formalism as can be found in
Ref.~\cite{book:RingSchuck1980}. Denoting the quasiparticle creation
operators by $b^\dagger_k$, we find that for a small fluctuation (labelled by
the two-quasi-particle index $kk'$)
around a given HFB state we
have to second order
\begin{eqnarray}
E&=&\bracket{}{\hat{H}}{}\quad,\\
H^{[20]}_{kk^\prime}&=&\bracket{}{b_{k^\prime}b_k\hat{H}}{}\quad,\\
\mathcal{A}_{kk^\prime ll^\prime} &=&\bracket{}{
b_{k^\prime}b_k\hat{H}b^\dagger_lb^\dagger_{l^\prime}}{}\quad,
\label{eqn:A}\\
\mathcal{B}_{kk^\prime ll^\prime}
&=&\bracket{}{
b_{k^\prime}b_k\hat{H}b_lb_{l^\prime}}{}\quad.
\label{eqn:B}
\end{eqnarray}
Here $E$ denotes the \ac{HB} vacuum energy, $H^{[20]}$ denotes the
gradient of the energy surface at that point and the matrices $\mathcal A$ and
$\mathcal B$ are the usual \ac{QRPA} matrices,
which are simple related to the mass and potential matrices of the 
\ac{LHE}.  The \ac{QRPA} equations may be
cast in the form \cite{book:RingSchuck1980}
\begin{eqnarray}
i(\mathcal{A}+\mathcal{B})P^\mu&=& Q^\mu\quad.\\
-i(\mathcal{A}-\mathcal{B})Q^\mu&=& \Omega_\mu^2P^\mu\quad.
\label{eqn:sub_lhe}
\end{eqnarray}
where $Q$ and $P$ are canonical coordinate and momentum variables,
respectively.  Furthermore from equations (\ref{eqn:sub_lhe}) we deduce
\begin{eqnarray}
(\mathcal{A}+\mathcal{B})(\mathcal{A}-\mathcal{B})Q^\mu=\Omega_\mu^2 Q^\mu\quad,
\label{eqn:lhe}
\end{eqnarray}
which is equivalent to the \ac{LHE} equation (\ref{eqn:local_qrpa}).

As sketched above, there are exact but spurious solutions to the
\ac{QRPA} which are particularly related zero eigenmodes of eigenvalue
equation (\ref{eqn:lhe}).  Operators associated with rotational
invariance in normal space $\hat{J}$ and in gauge space $\hat{N}$,
generate new states which resemble collective excitations, and thus
\begin{eqnarray}
\left[ \hat{H},\hat{N}\right ]&=&0\quad,\nonumber\\
\left[ \hat{H},\hat{J}_x\right ]=0\quad,%\nonumber\\
\left[ \hat{H},\hat{J}_y\right ]&=&-\omega i\hat{J}_z\quad,%\nonumber\\
\left[ \hat{H},\hat{J}_z\right ]=\omega i\hat{J}_y\quad.
\label{eqn:Jz}
\end{eqnarray}
The last two terms are zero for the $J=0$ ground state, but at finite $\omega$,
$iJ_y$ and $J_z$ form a canonical pair, a coordinate and a momentum.
In that case the set of spurious coordinates is
$\mathcal{F}^{s}=\{\mathcal{N},\mathcal{J}_x,\mathcal{J}_z\}$.

%\begin{table}%[tbhp]%h
%\caption[Spurious modes arising in the \ac{PPQ} model] {Spurious modes
%arising in the \ac{PPQ} model and their association with spontaneous
%symmetry breaking (for discussion see main text).}
%\label{table:spurious-ppq} 
%\begin{center}
%\begin{tabular}{l|c|c}
%%\ccline{1-7}
%\hline\hline
%Excitation 	&Deformation $\beta,~\gamma$	&Pairing $\Delta$\\
%\hline
%Symmetry Violation	&$[\hat{H},\hat{J}]=0$&$[\hat{H},\hat{N}]=0$\\
%\hline
%Associated Spurious Mode&$\mathcal{J}_x,~\mathcal{J}_z$&$\mathcal{N}$\\
%\hline\hline
%%\cline{1-7}
% \end{tabular} 
%\end{center} 
%\end{table}

%\input{tab/table-spurious-ppq}

\subsection{Results}

\begin{figure}%[tbhp]                     
\begin{center}                   
\includegraphics[clip,width=60mm]{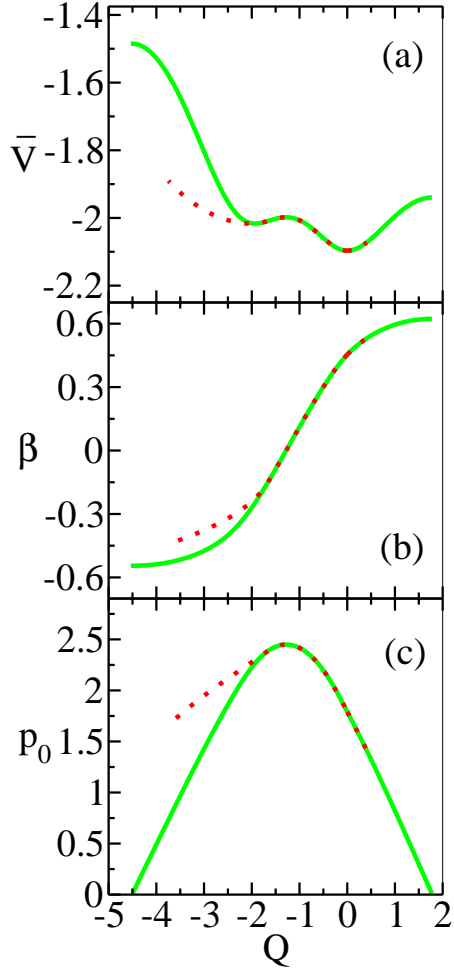}
\end{center}                
\caption[] {\label{fig:SA-E-beta-p0} \ac{ALACM} 
following the lowest QRPA mode as a function of the collective
coordinate $Q$.  Figure \ref{fig:SA-E-beta-p0}a shows the collective
potential $\bar{V}$, figure \ref{fig:SA-E-beta-p0}b shows the
$\beta$-deformation and figure \ref{fig:SA-E-beta-p0}c shows the
pairing parameter $p_0$.  Each panel shows the naive algorithm (dotted
line) and the zero-mode corrected one (solid line).  The scale of all
displayed quantities is arbitrary, for discussion of these see main
text.}
\end{figure}

We have first investigated the quality of the projection scheme for the
single $j$-shell \ac{PPQ} model.  First consider axially symmetric
($q_{\pm2}=0$) \ac{ALACM}, which corresponds to following the lowest
\ac{QRPA} mode as a function of the collective coordinate $Q$. The
model space is chosen to be small: $j=\frac{9}{2}$ and $N=4$ fermions
and with interaction strengths $\kappa=1$ and $\chi=0.333$.  We have
calculated the mean-field parameters along the collective path. The
results are shown in figure \ref{fig:SA-E-beta-p0}.

For the corrected algorithm (solid line) we identify two minima and
three maxima at $Q\sim\{-2,0\}$ and $Q\sim\{-5,-1.25,2\}$,
respectively. The lowest-energy solution at the starting
point $Q=0$ corresponds to a prolate minimum and the second minimum at
$Q\sim-2$ is oblate. Comparison of the collective potential
with the deformation parameter $\beta$ and pairing parameter $p_0$
exhibits the well-known competition between pairing and quadrupole
forces. At the boundaries of the collective coordinate $Q\sim\{-5,2\}$
the system shows maximal oblate and prolate deformation, respectively,
but the pairing field collapses, $p_0\sim0$. The converse is true at
the central maximum where the system has collapsed into a state of
pure-pairing; $q_{\pm M}\sim0$.

\begin{figure}%[tbhp]                     
\begin{center}                   
\includegraphics[clip,width=120mm]{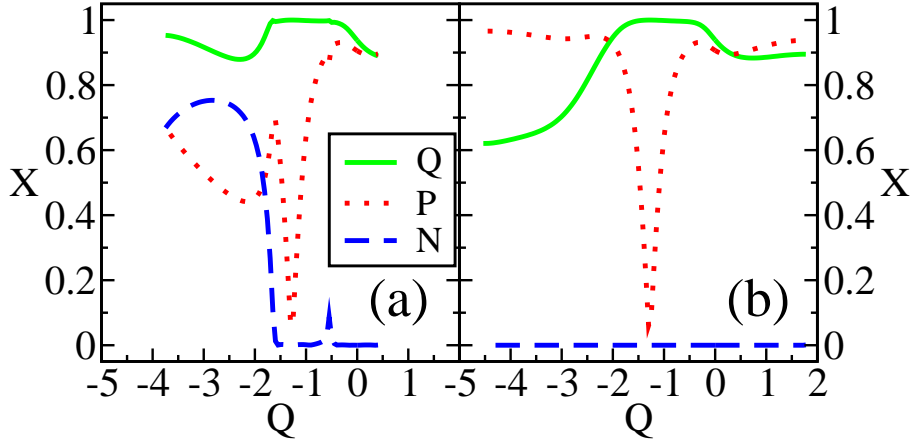}% 
\end{center}                
\caption[]{\label{fig:SA-CC}The overlap of the collective operator
with the quadrupole, pairing, and particle-number operators as a
function of the collective coordinate $Q$.  The solid line shows
$X_Q$, the dotted line $X_P$, and the dashed line $X_N$,
Eq.~(\ref{eqn:measures_overlaps}). Figure \ref{fig:SA-CC}a displays
the results for the uncorrected algorithm and figure \ref{fig:SA-CC}b
for the corrected algorithm.}
\end{figure}

Figure \ref{fig:SA-E-beta-p0} shows clearly that 
without correctly removing spurious modes
 (dotted line) the 
\ac{ALACM} algorithm fails outside  a limited range
of the collective coordinate $-3.75\lesssim Q\lesssim0.5$; we note
that the collective potential and associated mean-field parameters
follow a different path from the corrected algorithm (solid line).  To
better understand this we turn to an analysis of the collective
coordinate and calculate the degree of overlap of the collective
operator with the operators contained in the Hamiltonian:
$\hat{O}\in\{\hat{Q}_0,(\hat{P}^\dagger+\hat{P}),\hat{N}\}$,
%We obtain
%the quantities
%\begin{subequations}
%\label{eqn:measures_overlaps}
%\begin{eqnarray}
%X_Q&=&\sum_\alpha\mathcal{F}_{,\alpha}\mathcal{Q}_{0,\alpha}\quad,\\
%X_P&=&\sum_\alpha\mathcal{F}_{,\alpha}(\mathcal{P}^\dagger
%+\mathcal{P})_{,\alpha}\quad,\\
%X_N&=&\sum_\alpha\mathcal{F}_{,\alpha}\mathcal{N}_{,\alpha}\quad,
%%\label{eqn:static_likeness}
%\end{eqnarray}
%\end{subequations}
%where the normalisation condition $\sum_\alpha\mathcal{O}_{,\alpha}=1$
%holds $\forall \mathcal{O}\in\{\mathcal{F},\mathcal{Q}_0,
%(\mathcal{P}^\dagger+\mathcal{P}),\mathcal{N}\}$. 
\begin{equation}
X_O=\sum_\alpha \mathcal{F}_{,\alpha}\mathcal{O}_{\alpha}\quad.
\label{eqn:measures_overlaps}
\end{equation}
The results are shown in figure \ref{fig:SA-CC}.

For the naive algorithm shown in figure \ref{fig:SA-CC}a  ({\it c.f.}  figure
\ref{fig:SA-E-beta-p0}c) we see that the collective coordinate is dominated
in the region of maximal pairing by a quadrupole-like operator
($\mathcal{F}\sim{Q}_0$). In particular, where $Q\sim-1.25$
the pairing-like parameter $X_P$ tends to zero, and we have a pure
$\beta$-vibration.  From figure \ref{fig:SA-CC}a we see clearly why
the naive algorithm fails at $Q\sim-1.5$; first, we see a sudden
increase in the parameter $X_N$, and secondly, a sharp change in the
pairing-like parameter $X_P$.  From this it is evident that the the
spurious coordinate associated with particle number $\mathcal{N}$ is
no longer sufficiently decoupled from the collective coordinate.

\begin{figure}%[tbhp]                     
\begin{center}                   
\includegraphics[clip,width=80mm]{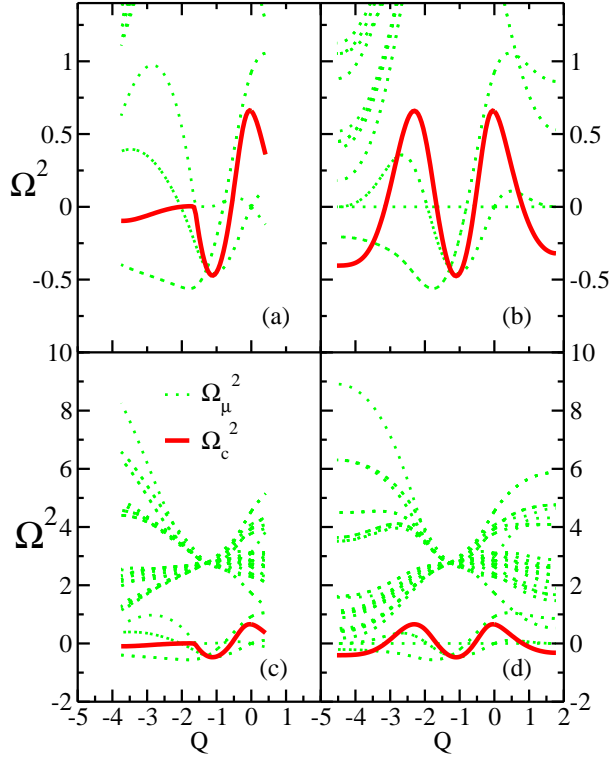}
\end{center}                
\caption[]{\label{fig:SA-QRPA}QRPPA energy eigenvalues squared (dotted
lines).  The solid line shows the QRPA mode selected by the path
following algorithm.
Figures \ref{fig:SA-QRPA}a and \ref{fig:SA-QRPA}c display results for the
uncorrected algorithm and figures \ref{fig:SA-QRPA}b and
\ref{fig:SA-QRPA}d display corresponding results for the corrected algorithm.  The scale of all
displayed quantities is arbitrary, and determined by the value of
$\kappa=1$.}
\end{figure} 

\begin{figure}%[tbhp]                     
\begin{center}                   
\includegraphics[clip,width=80mm]{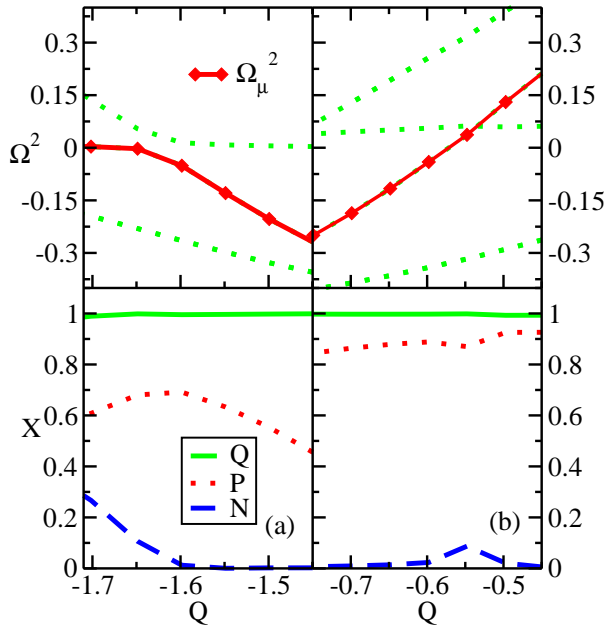}
\end{center}                
\caption[]
{\label{fig:SA-CC-zoom}Analysis of the
eigenvalues of the QRPA. Top boxes shows eigenvalues of the QRPA
(dotted lines) following the lowest QRPA mode (solid line) as a
function of the collective coordinate $Q$. Diamond symbols
indicate computed points (joined by linear interpolation). Bottom box
shows the measure of the overlaps $X_O$.}
\end{figure}

This can be seen in figure
\ref{fig:SA-QRPA}, where we study $\Omega^2$, Eq.~(\ref{eqn:lhe}).
On the left we show the results of the naive algorithm, and on the
right the results from the zero-mode corrected one; the upper two
figures differ from the lower two only in scale. It can be seen from figure
\ref{fig:SA-QRPA}a that at $Q\sim-1.5$ the energy of the selected
eigenmode changes abruptly. At this point we note that the eigenvalue
of the collective mode $\Omega^2_c$ and that of the mode associated
with the particle number operator $\Omega^2_N$ have become degenerate
and an avoided crossing of the two levels has occurred. Here the path
following algorithm selects the wrong eigenmode, {\it i.e.} a spurious
mode rather than the collective mode, leading the system away from the
collective path --- this is seen most clearly in figure
\ref{fig:SA-CC-zoom}. In figure \ref{fig:SA-QRPA}b is shown the
corrected result whereby the collective coordinate smoothly crosses
the level associated with particle number.

In order to illustrate the source of the problems in 
the naive algorithm we study both the real and avoided
crossings present in figure \ref{fig:SA-QRPA}a.
We see that the avoided crossing has a large admixture
of particle number, and that even the real crossing has a smallish
admixture (indicating it is probably a very narrow avoided crossing).
We have further investigated the nature of the avoided crossing
by using a variety of step lengths. We find that the result
is not influenced by changing the step size, and that we can
not take steps so large that we run straight through the avoided crossing.

\begin{figure}%[tbhp]                     
\begin{center}                   
\includegraphics[clip,width=50mm]{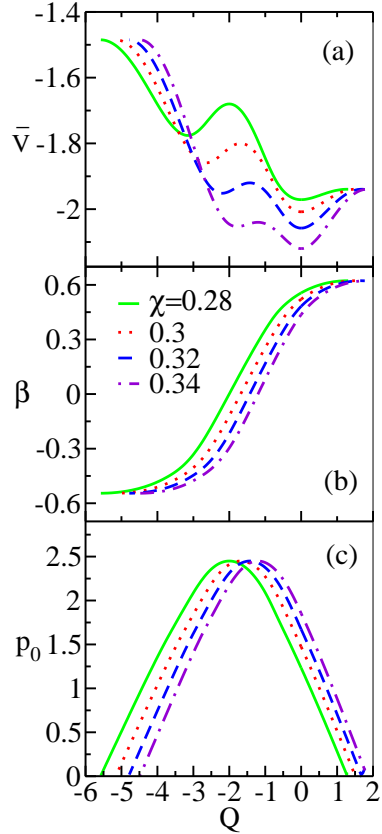}
\end{center}                
\caption[Comparison of ALACM for varying pairing strengths.]
{\label{fig:SA-proj-chi-all} The result of varying the pairing strength.
See  figure \ref{fig:SA-E-beta-p0} for an explanation of the subgraphs.
 }
\end{figure}

We have investigated changes in the collective potential with respect
to variations of the pairing strength $\chi$ from $0.28$ to
$0.34$. The results are shown in figure
\ref{fig:SA-proj-chi-all}.  From figure \ref{fig:SA-proj-chi-all}a we
see that the collective potential is higher in energy for lower
pairing strength and with a higher maximum relative to the
ground-state. Both ends of the collective potential terminate at the
same energy, since each of the five systems are identical, that is,
where only the quadrupole moment is in effect yielding maximally
deformed axially symmetric spheroids.  Moreover, we observe from
figures \ref{fig:SA-proj-chi-all}b-\ref{fig:SA-proj-chi-all}c that the
maximum pairing and $\beta$-deformation attained by the system is
identical in all cases of pairing strength $\chi$ as is expected.  

%The startling feature of this comparison is that at the end points of
%the collective path yields a shorter potential for higher pairing
%strength.  We would expect that for weaker pairing strength the
%deformation parameter dominates quicker causing an earlier pairing
%collapse. As we have just stated, the converse is true.  The reason
%that pairing collapse is reached sooner for strong pairing is, we
%believe, that a stronger pairing strength results in a lower maximum
%thus allowing the system to pass through the spherical maximum and
%into pairing collapse sooner, resulting in a shorter collective path.

We conclude that the zero-mode-corrected algorithm seems to be able to
deal much more efficiently with crossing modes than the naive one. The
results obtained with this algorithm have all the properties of
the full results, and do not suffer from spurious admixtures and
avoided crossings. Since finally we want to understand what happens 
for rotating nuclei, we must look further at collective motion
at non-zero angular momentum.

\section{PPQ Model at finite rotational velocity}
\label{sec:single_j_ppq_model-frv}

\begin{figure}%[tbhp]
\begin{center}
\includegraphics[clip,height=80mm]{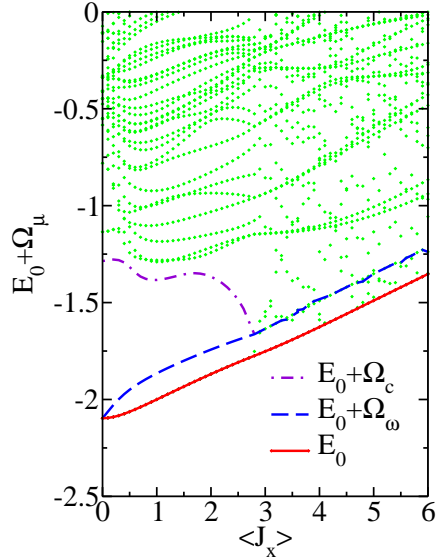}
\caption[Partial energy spectra of the PPQ model at finite rotational velocity]
{\label{fig:energy-levels-ppq-rot} Partial energy spectra of the PPQ
model at finite rotational velocity.  Figure shows ground-state energy
and a selection of QRPA eigenvalues for fixed quadrupole strength
$\kappa=10$, fixed pairing strength, $\chi=0.333$, and rotational
velocity $0\leq J_x\leq6$.  The scale of all displayed quantities is
arbitrary.}
\end{center}
\end{figure}

To perform a representative test-case of \ac{ALACM} at finite
rotational velocity we have minimised the mean-field energy, and
computed the \ac{QRPA} eigenvalues for values of angular momentum
$0\leq J_x\leq6$, where $J_x=\expectation{\hat{J}_x}$.  The results
are shown in figure \ref{fig:energy-levels-ppq-rot}.

Figure \ref{fig:energy-levels-ppq-rot} shows 
the behaviour of the ground state and its \ac{QRPA} excitations
$E_\mu=E_0+\Omega_\mu$, where $E_0$ is the ground-state energy and
$\Omega_\mu$ is the $\mu$th 
\ac{QRPA} frequency.  
As the angular momentum increases, the pairing field weakens and
collapses to zero for $J_x=3$. Until that collapse
the ground-state (solid
line) is degenerate with two eigenmodes associated with the two spurious
operators $\mathcal{N}$ and $\mathcal{J}_x$. 
Thereafter, the only spurious operator is $J_x$.
The dashed line corresponds to a solution of the
\ac{QRPA} $\mathcal{F}^\omega$ that appears at the value of the
rotational frequency squared ($\omega^2_x$), corresponding to 
the conjugate pair of operators $J_y$ and $J_z$.
 Most
interesting here, is the appearance of the lowest-lying excitation
corresponding to a collective mode $\mathcal{F}^c$ (dot-dashed
line). As the rotational energy increases the collective excitation
comes closer to the ground-state.

Until pairing collapse modes mix between pairing and deformation
modes.  At the point of collapse such modes decouple and we should
really solve the ordinary RPA. Nevertheless, even solving QRPA we can
easily distinguish between the excitations that change particle number
and the deformational modes (continuing lines) which continue to be
well behaved into the realm of no pairing. These modes correspond to
excitation levels of the standard Nilsson model
\cite{book:RingSchuck1980}.

\subsection{Model results at finite rotational velocity}

We study 
angular momentum $J_x=2$, such that the ground-state of the system has
non-zero pairing and quadrupole effects ({\it c.f.} figure
\ref{fig:energy-levels-ppq-rot}), we have computed \ac{ALACM} for the
single $j$-shell \ac{PPQ} model at finite rotational velocity
following the lowest collective mode of the \ac{QRPA}. For
illustrative purposes we will in the first instance remove only
admixtures with the particle number operator. 

%In order to analyse the collective coordinate in detail we consider
%the measures of overlap:
%\begin{subequations}
%\label{eqn:rot_overlap}
%\begin{eqnarray}
%X_{O}=\sum_\alpha\mathcal{F}_{,\alpha}\mathcal{O}_{,\alpha}\quad.
%\end{eqnarray}
%\end{subequations}
%Here
%$\mathcal{O}_{,\alpha}\in\{\mathcal{F}_{,\alpha},\mathcal{Q}_{0,\alpha},
%(\mathcal{P}^\dagger+\mathcal{P})_{,\alpha},\mathcal{N}_{,\alpha}
%,\mathcal{J}_{x,\alpha},\mathcal{J}_{z,\alpha}\}$. 
The quantities $X_O\in\{X_Q, X_P, X_N,X_{J_x},X_{J_z}\}$ give the measures
of overlap for the collective operator with the quadrupole, pairing,
particle number and rotational operators, respectively.

\begin{figure}%[tbhp]                     
\begin{center}                   
\begin{tabular}{cc}           
\includegraphics[clip,width=80mm]{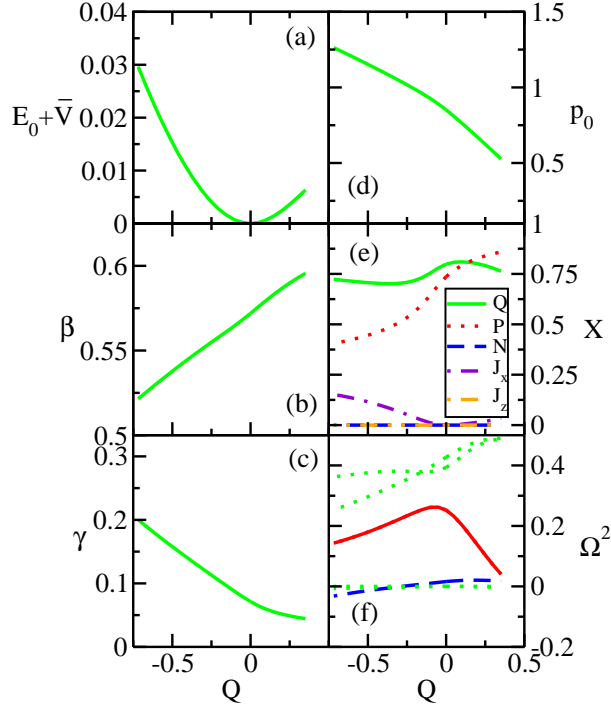}
\end{tabular}
\end{center}                
\caption[ALACM for the PPQ model at finite rotational velocity]
{\label{fig:rot-alacm-proj-all-unstable} ALACM for the PPQ model at
finite rotational velocity following the lowest QRPA mode as a
function of the collective coordinate $Q$. The scale of all displayed
quantities is arbitrary.}
\end{figure}

The results are shown in figure \ref{fig:rot-alacm-proj-all-unstable}.
As can be seen, the computation has produced values for $-0.6<Q<0.3$
only, due to an instability arising in the path following algorithm
discussed further below. For negative values of the collective
coordinate we see a decrease in $\beta$-deformation
(\ref{fig:rot-alacm-proj-all-unstable}b) and an increase in
$\gamma$-deformation (\ref{fig:rot-alacm-proj-all-unstable}c),
indicating that the system is becoming more triaxial, while the
pairing parameter $p_0$ increases as well
(\ref{fig:rot-alacm-proj-all-unstable}d).  We thus expect the
components of the $M=\pm2$ quadrupole operator to be important.
% since it is these components that ultimately lead to triaxial deformation.
%This possibility, as well as that of $M=\pm1$
%quadrupole components is considered in detail further on in the
%investigation.

As can be seen from figure \ref{fig:rot-alacm-proj-all-unstable}e
a spurious admixture
connected with $\mathcal{J}_x$ is causing problems for $Q<0$.
As can be seen from figure
\ref{fig:rot-alacm-proj-all-unstable}f the eigenvalue of the
collective coordinate $\Omega_c$ (solid line) and the eigenvalue of
the coordinate associated with the spurious mode $\Omega_\omega$
(dashed line) are approximately degenerate. There are therefore two
approximately degenerate paths for the path-following algorithm to
distinguish, one of which is a spurious solution.

We note one interesting factor outstanding from our discussion of the
results thus far.  It can be seen that the collective coordinate
represented in figure \ref{fig:rot-alacm-proj-all-unstable}f runs
approximately parallel with the next higher lying state and though we
are unsure of the physical significance of this, it would be
interesting to analysis the collective and non-collective coordinates
in more detail. To this end, better technology aimed at giving a more
detailed description of the system will be introduced below.

We now project out all components associated with spurious modes
$\mathcal{N}~,\mathcal{J}_x$, and $\mathcal{J}_z$, and secondly, we
will compute the measures of overlap associated with all operators of
the \ac{PPQ} model.

\begin{figure}%[tbhp]                     
\begin{center}                
\begin{tabular}{cc}           
\includegraphics[clip,width=80mm]{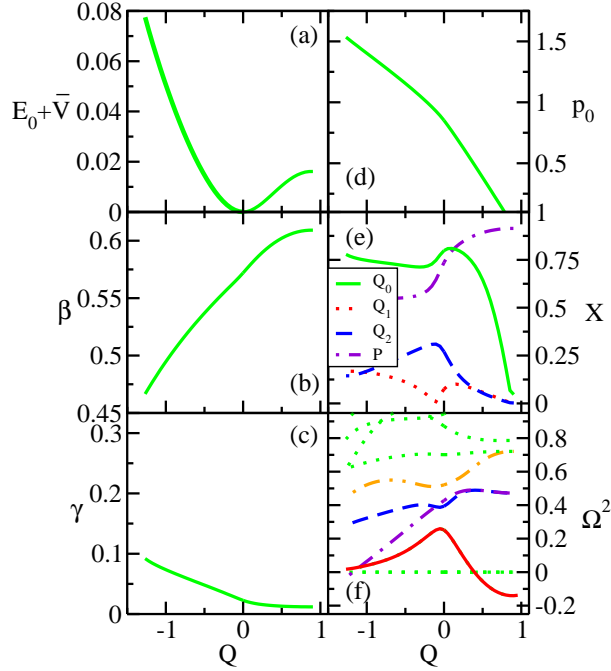}
\end{tabular}
\end{center}                
\caption[ALACM for the PPQ model at finite rotational velocity ---
generalised alternative] {\label{fig:rot-alacm-projJxJz-all-unstable}
ALACM for the PPQ model at finite rotational velocity following the
lowest QRPA mode as a function of the collective coordinate.  The
scale of all displayed quantities is arbitrary.}
\end{figure}

\begin{figure}%[tbhp]                     
\begin{center}                
\begin{tabular}{c}           
\includegraphics[clip,width=45mm]{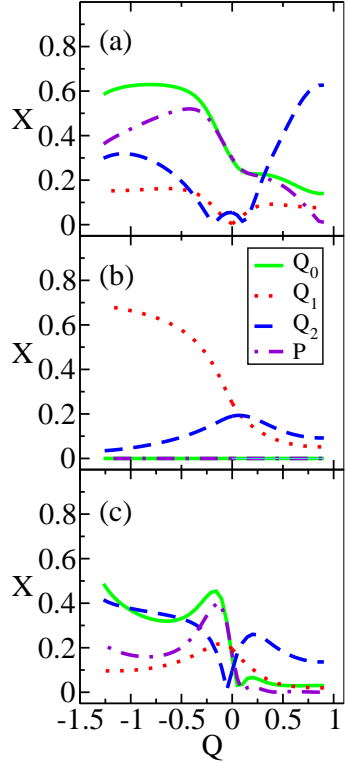}
\end{tabular}
\end{center}                
\caption[Analysis of low-lying eigenmodes of the QRPA]
{\label{fig:rot-alacm-projJxJz-QRPA-x-unstable} Analysis of low-lying
eigenmodes of the QRPA eigenvalues of the QRPA following the lowest QRPA
mode as a function of the collective coordinate.  The scale of all
displayed quantities is arbitrary. (For discussion of displayed
quantities see main text.)}
\end{figure}

We work with the four quantities $\mathcal{O}\in\{\mathcal{Q}_0,
(\mathcal{Q}_{+1}+\mathcal{Q}_{-1}),
(\mathcal{Q}_{+2}+\mathcal{Q}_{-2}), (\mathcal{P}^\dagger+\mathcal{P})
\}$.  Here on, $X_0\in\{X_{J_x},X_{J_z},X_N\}$ --- the quantities
connected with projection --- are zero eigenvalues of the
\ac{QRPA} eigenvalue equation and otherwise will not be alluded
to. The results are shown in figures
\ref{fig:rot-alacm-projJxJz-all-unstable}-\ref{fig:rot-alacm-projJxJz-QRPA-x-unstable}.

%Here, we work with the four quantities
%\begin{subequations}
%\begin{eqnarray}
%X_{Q_0}&=&\sum_\alpha\mathcal{F}_{,\alpha}\mathcal{Q}_{0,\alpha}\quad,\label{eqn:Q0}\\
%X_{Q_1}&=&\sum_\alpha\mathcal{F}_{,\alpha}(\mathcal{Q}_{-1,\alpha}+\mathcal{Q}_{1,\alpha})\quad,\label{eqn:Q1}\\
%X_{Q_2}&=&\sum_\alpha\mathcal{F}_{,\alpha}(\mathcal{Q}_{-2,\alpha}+\mathcal{Q}_{2,\alpha})\quad,\label{eqn:Q2}\\
%X_{P}
%&=&\sum_\alpha\mathcal{F}_{,\alpha}(\mathcal{P}^\dagger_{,\alpha}+\mathcal{P}_{,\alpha})\quad,
%\label{eqn:P}
%\end{eqnarray}
%\end{subequations}
%whose meaning is to measure the overlap of the collective coordinate
%$\mathcal{F}^c$ with the quadrupole operator
%(\ref{eqn:Q0}-\ref{eqn:Q2}) and the pairing operator (\ref{eqn:P}).
%Here on, $X_0\in\{X_{J_x},X_{J_z},X_N\}$ (the quantities connected
%with projection) are related to zero eigenvalues of the \ac{QRPA}
%eigenvalue equation and will otherwise be not alluded to. The results
%are shown in figures
%\ref{fig:rot-alacm-projJxJz-all-unstable}-\ref{fig:rot-alacm-projJxJz-QRPA-x-unstable}.

{}From figure \ref{fig:rot-alacm-projJxJz-all-unstable} we see that for
positive values of the collective coordinate the collective path has
been successfully computed to its end. This is identified by pairing
collapse $p_0\sim0$ for  $Q\sim0.8$
(figure \ref{fig:rot-alacm-projJxJz-all-unstable}d) at a
maximum in the collective potential (figure
\ref{fig:rot-alacm-projJxJz-all-unstable}a).  As the system approaches
pairing collapse the $\beta$-deformation approaches its maximum value
and the $\gamma$-deformation approaches a minimal value (figure
\ref{fig:rot-alacm-projJxJz-all-unstable}b). In other words, at
pairing collapse the system is maximally axially deformed with some
small admixture of triaxial deformation. For $0\leq Q\lesssim0.8$ the
collective coordinate becomes increasingly more pairing-like for
greater values of $Q$ and less quadrupole-like. In particular, this
statement is true of \emph{all} components of the quadrupole operator,
{\it i.e.} for $X_{Q_0},~X_{Q_1}$ and $X_{Q_2}$, (solid, dotted and
dashed lines respectively) and thus the collective coordinate becomes
very pairing-like trying to force the system back toward equilibrium.

For negative values of the collective coordinate the results show an
instability at $Q\sim-1.25$, where the computation has failed. From figure
\ref{fig:rot-alacm-projJxJz-all-unstable}e we see 
that the collective coordinate $\Omega_c$ (solid line), becomes 
degenerate with a second coordinate, labelled $\Omega_6$ (dashed
line). Thus the assumption of a single collective coordinate fails here.
This is quite common in this approach, and requires
and extended algorithm.

In order to convince ourselves that this is the correct physics, and
we are not looking at the spurious admixture problem again, we analyse
the \ac{QRPA} modes $\Omega_\mu$ in some detail. We have computed the
measures of overlap for a few of these modes. At equilibrium there are
three spurious modes, and the fourth eigenmode (next highest in
energy) is the collective eigenmode we are following $\Omega_c$. We
concentrate our efforts on the next few \ac{QRPA} modes, namely, the
fifth $\Omega_5$ (dashed line), sixth $\Omega_6$ (dot-dashed line),
and the seventh $\Omega_7$ (dot-dot-dashed line, in figure
\ref{fig:rot-alacm-projJxJz-all-unstable}d).  We note from figure
\ref{fig:rot-alacm-projJxJz-all-unstable}f that these modes do not
remain in this order for all values of the collective coordinate since
there are crossings of energy levels in the energy spectra. The
labelling of these modes however, shall remain fixed as defined at
$Q=0$, regardless of their subsequent order. The results are shown in
figure \ref{fig:rot-alacm-projJxJz-QRPA-x-unstable}.

We have computed the measure of overlap 
\begin{eqnarray}
X_O=\sum_\alpha\mathcal{F}^\mu_{,\alpha}\mathcal{O}_{,\alpha}\quad,
&&\forall\mu\in\{5,6,7\}\quad.
\end{eqnarray}
Figure \ref{fig:rot-alacm-projJxJz-QRPA-x-unstable} shows the results
of our analysis for $\Omega_5$ (figure
\ref{fig:rot-alacm-projJxJz-QRPA-x-unstable}a), $\Omega_6$ (figure
\ref{fig:rot-alacm-projJxJz-QRPA-x-unstable}b), and $\Omega_7$ (figure
\ref{fig:rot-alacm-projJxJz-QRPA-x-unstable}c).  In each sub-figure is
shown the measures of overlap $X_{Q_0}$ (solid line), $X_{Q_1}$
(dotted line), $X_{Q_2}$ (dashed line), and $X_{P}$ (dot-dashed
line). 

The fifth eigenvalue (figure
\ref{fig:rot-alacm-projJxJz-QRPA-x-unstable}c) and seventh eigenvalue
(figure \ref{fig:rot-alacm-projJxJz-QRPA-x-unstable}a) both exhibit a
complex arrangement of all measures of overlap, whose evolutions are
not especially discernible; we will return to this later. The sixth
eigenmode, figure \ref{fig:rot-alacm-projJxJz-QRPA-x-unstable}b,
contains exclusively admixtures of $X_{Q_1}$ and $X_{Q_2}$ meaning
that this is a mode that drives both $\gamma$-deformation
($\hat{Q}_{\pm2}$) and tilting ($\hat{Q}_{\pm1}$). 
At the point of instability, $Q\sim -1.25$, the
tilting
character is particularly strong.

\section{Summary and outlook}
\label{sec:summary_and_outlook_single_jshell}

In this article we have investigated \ac{ALACM} using the \ac{LHA} for
the single $j$-shell \ac{PPQ} model. We have computed the collective
path, as well as a variety of characteristics of the mean-field as a
function of a single collective coordinate.  In order to analyse the
collective coordinate in more detail we introduced the measures of
overlap, which turned out to be an indispensable tool to analyse the
collective coordinate especially in regions where energy levels become
degenerate. We introduced a projection technique to remove spurious
admixture, which works by projecting a subspace out of the computed
mass matrix and potential matrix free from spurious admixture. This
method is easily extended to remove the admixture of any finite number
of spurious modes.  The mixing of spurious modes with the collective
mode brings to light an inadequacy of our initial assumption that all
modes are approximately decoupled from the collective coordinate.

Using the \ac{LHA}, collective motion was shown to be non-linear in
nature and dominated by neither pairing nor quadrupole degrees of
freedom exclusively. Rather, the system self-selects either pairing,
or quadrupole, or an admixture of the two to dominate the dynamics and
thus select how best the system deforms.  More especially, we have
found the remarkable feature of a shorter collective path for stronger
pairing strength which is not an obvious result.

This leads us to an important limitation, namely, that a single
collective coordinate is not sufficient to describe \ac{ALACM} in
regions where the collective coordinate undergoes real crossings with
other physical solutions which mix with the collective coordinate.
Furthermore, in the course of this investigation we have stumbled
across the apparent importance of tilting solutions arising in the
quadrupole moment.

We have not computed the covariant derivative of the potential, but we
have rather computed a truncation of this which leaves a normal
derivative and losing some terms. These terms may well be more
important than we suspected at the start. This is something we would
like to investigate further.  In addition to this, we have attempted
to work within the full configuration space available in order to
allow a large set of degrees of freedom to contribute to the
dynamics. It is worth mentioning here that work on a semi-realistic
approach to \ac{ALACM} using the \ac{PPQ} model and incorporating many
$j$-shells \cite{article:AlmehedWalet2003} uses a truncated \ac{QRPA}
which inadvertently avoids handling spurious modes directly. The
projected \ac{QRPA} was introduced as a computational aid where it was
assumed that coupling of normal and spurious modes with the collective
coordinate would be small. As we have shown, this assumption may not
be correct.

Finally, we note that the \ac{LHA} is not restricted to a solution for
a collective path only, {\it i.e.} we could take into account more
than one collective coordinate to describe the dynamics of
\ac{ALACM}. The next step is to modify the
algorithms we have built, such that they are more apt to describe
nuclei at finite rotational velocity using two collective coordinates
to determine a collective surface.  
%In this way we may be able to
%surmount the problems found in this investigation which, up to now,
%are unresolved.  This would be both a new and an interesting
%development to the theory and will be made a course of investigation
%in the near future.

\subsection*{Acknowledgements}

This research was funded by the EPSRC and a UMIST scholarship scheme.

%%%%%% \bibliography{physics-articles.bib,physics-books.bib}

\begin{thebibliography}{21}
\expandafter\ifx\csname natexlab\endcsname\relax\def\natexlab#1{#1}\fi
\expandafter\ifx\csname bibnamefont\endcsname\relax
  \def\bibnamefont#1{#1}\fi
\expandafter\ifx\csname bibfnamefont\endcsname\relax
  \def\bibfnamefont#1{#1}\fi
\expandafter\ifx\csname citenamefont\endcsname\relax
  \def\citenamefont#1{#1}\fi
\expandafter\ifx\csname url\endcsname\relax
  \def\url#1{\texttt{#1}}\fi
\expandafter\ifx\csname urlprefix\endcsname\relax\def\urlprefix{URL }\fi
\providecommand{\bibinfo}[2]{#2}
\providecommand{\eprint}[2][]{\url{#2}}

\bibitem[{\citenamefont{Blaizot and Ripka}(1986)}]{book:BlaizotRipka1986}
\bibinfo{author}{\bibfnamefont{J.~P.} \bibnamefont{Blaizot}} \bibnamefont{and}
  \bibinfo{author}{\bibfnamefont{G.}~\bibnamefont{Ripka}},
  \emph{\bibinfo{title}{Quantum Theory of Finite Systems}}
  (\bibinfo{publisher}{MIT Press}, \bibinfo{address}{Berlin},
  \bibinfo{year}{1986}).

\bibitem[{\citenamefont{Ring and Schuck}(1980)}]{book:RingSchuck1980}
\bibinfo{author}{\bibfnamefont{P.}~\bibnamefont{Ring}} \bibnamefont{and}
  \bibinfo{author}{\bibfnamefont{P.}~\bibnamefont{Schuck}},
  \emph{\bibinfo{title}{The Nuclear Many-Body Problem}}
  (\bibinfo{publisher}{Springer-Verlag}, \bibinfo{address}{Berlin},
  \bibinfo{year}{1980}).

\bibitem[{\citenamefont{Bohr and Mottleson}(1955)}]{article:BohrMottleson1955}
\bibinfo{author}{\bibfnamefont{A.}~\bibnamefont{Bohr}} \bibnamefont{and}
  \bibinfo{author}{\bibfnamefont{B.}~\bibnamefont{Mottleson}},
  \bibinfo{journal}{Mat. Fys. Medd. Dan. Vid. Selsk}
  \textbf{\bibinfo{volume}{30}}, \bibinfo{pages}{1} (\bibinfo{year}{1955}).

\bibitem[{\citenamefont{Baranger and Kumar}(1965)}]{article:BarangerKumar1965}
\bibinfo{author}{\bibfnamefont{M.}~\bibnamefont{Baranger}} \bibnamefont{and}
  \bibinfo{author}{\bibfnamefont{K.}~\bibnamefont{Kumar}},
  \bibinfo{journal}{Nucl. Phys.} \textbf{\bibinfo{volume}{62}},
  \bibinfo{pages}{113} (\bibinfo{year}{1965}).

\bibitem[{\citenamefont{Baranger and
  Kumar}(1968{\natexlab{a}})}]{article:BarangerKumar1968}
\bibinfo{author}{\bibfnamefont{M.}~\bibnamefont{Baranger}} \bibnamefont{and}
  \bibinfo{author}{\bibfnamefont{K.}~\bibnamefont{Kumar}},
  \bibinfo{journal}{Nucl. Phys. A} \textbf{\bibinfo{volume}{110}},
  \bibinfo{pages}{490} (\bibinfo{year}{1968}{\natexlab{a}}).

\bibitem[{\citenamefont{Baranger and
  Kumar}(1968{\natexlab{b}})}]{article:BarangerKumar1968b}
\bibinfo{author}{\bibfnamefont{M.}~\bibnamefont{Baranger}} \bibnamefont{and}
  \bibinfo{author}{\bibfnamefont{K.}~\bibnamefont{Kumar}},
  \bibinfo{journal}{Nucl. Phys. A} \textbf{\bibinfo{volume}{110}},
  \bibinfo{pages}{529} (\bibinfo{year}{1968}{\natexlab{b}}).

\bibitem[{\citenamefont{Baranger and
  Kumar}(1968{\natexlab{c}})}]{article:BarangerKumar1968c}
\bibinfo{author}{\bibfnamefont{M.}~\bibnamefont{Baranger}} \bibnamefont{and}
  \bibinfo{author}{\bibfnamefont{K.}~\bibnamefont{Kumar}},
  \bibinfo{journal}{Nucl. Phys. A} \textbf{\bibinfo{volume}{112}},
  \bibinfo{pages}{241} (\bibinfo{year}{1968}{\natexlab{c}}).

\bibitem[{\citenamefont{Baranger and
  Kumar}(1968{\natexlab{d}})}]{article:BarangerKumar1968d}
\bibinfo{author}{\bibfnamefont{M.}~\bibnamefont{Baranger}} \bibnamefont{and}
  \bibinfo{author}{\bibfnamefont{K.}~\bibnamefont{Kumar}},
  \bibinfo{journal}{Nucl. Phys. A} \textbf{\bibinfo{volume}{112}},
  \bibinfo{pages}{273} (\bibinfo{year}{1968}{\natexlab{d}}).

\bibitem[{\citenamefont{Dang et~al.}(2000)\citenamefont{Dang, Klein, and
  Walet}}]{review:DangKleinWalet2000}
\bibinfo{author}{\bibfnamefont{G.~D.} \bibnamefont{Dang}},
  \bibinfo{author}{\bibfnamefont{A.}~\bibnamefont{Klein}}, \bibnamefont{and}
  \bibinfo{author}{\bibfnamefont{N.~R.} \bibnamefont{Walet}},
  \bibinfo{journal}{Phys. Rep.} \textbf{\bibinfo{volume}{335}},
  \bibinfo{pages}{93} (\bibinfo{year}{2000}).

\bibitem[{\citenamefont{Nakatsukasa and
  Walet}(1998{\natexlab{a}})}]{article:NakatsukasaWalet1998c}
\bibinfo{author}{\bibfnamefont{T.}~\bibnamefont{Nakatsukasa}} \bibnamefont{and}
  \bibinfo{author}{\bibfnamefont{N.~R.} \bibnamefont{Walet}},
  \bibinfo{journal}{Czech. J. Phys.} \textbf{\bibinfo{volume}{48}},
  \bibinfo{pages}{813} (\bibinfo{year}{1998}{\natexlab{a}}).

\bibitem[{\citenamefont{Nakatsukasa and
  Walet}(1998{\natexlab{b}})}]{article:NakatskasaWalet1998}
\bibinfo{author}{\bibfnamefont{T.}~\bibnamefont{Nakatsukasa}} \bibnamefont{and}
  \bibinfo{author}{\bibfnamefont{N.~R.} \bibnamefont{Walet}},
  \bibinfo{journal}{Phys. Rev. C} \textbf{\bibinfo{volume}{57}},
  \bibinfo{pages}{1192} (\bibinfo{year}{1998}{\natexlab{b}}).

\bibitem[{\citenamefont{Nakatsukasa and
  Walet}(1998{\natexlab{c}})}]{article:NakatskasaWalet1998b}
\bibinfo{author}{\bibfnamefont{T.}~\bibnamefont{Nakatsukasa}} \bibnamefont{and}
  \bibinfo{author}{\bibfnamefont{N.~R.} \bibnamefont{Walet}},
  \bibinfo{journal}{Phys. Rev. C} \textbf{\bibinfo{volume}{58}},
  \bibinfo{pages}{3397} (\bibinfo{year}{1998}{\natexlab{c}}).

\bibitem[{\citenamefont{Kobayasi et~al.}(2000)\citenamefont{Kobayasi,
  Nakatsukasa, Matsuo, and
  Matsuyanagi}}]{article:KobayasiNakatsukasaMatsuoMatsuyanagi2000}
\bibinfo{author}{\bibfnamefont{M.}~\bibnamefont{Kobayasi}},
  \bibinfo{author}{\bibfnamefont{T.}~\bibnamefont{Nakatsukasa}},
  \bibinfo{author}{\bibfnamefont{M.}~\bibnamefont{Matsuo}}, \bibnamefont{and}
  \bibinfo{author}{\bibfnamefont{K.}~\bibnamefont{Matsuyanagi}},
  \bibinfo{journal}{arXiv:nucl-th} \textbf{\bibinfo{volume}{0001056}},
  \bibinfo{pages}{\,} (\bibinfo{year}{2000}).


\bibitem[{\citenamefont{Almehed and Walet}(2004)}]{article:AlmehedWalet2003}
\bibinfo{author}{\bibfnamefont{D.}~\bibnamefont{Almehed}} \bibnamefont{and}
  \bibinfo{author}{\bibfnamefont{N.~R.} \bibnamefont{Walet}},
  \bibinfo{journal}{Phys. Rev. C} \textbf{\bibinfo{volume}{69}},
  \bibinfo{pages}{024302} (\bibinfo{year}{2004}).

\bibitem[{\citenamefont{Almehed and Walet}(2004)}]{article:AlmehedWalet2004}
\bibinfo{author}{\bibfnamefont{D.}~\bibnamefont{Almehed}} \bibnamefont{and}
  \bibinfo{author}{\bibfnamefont{N.~R.} \bibnamefont{Walet}},
  \bibinfo{journal}{Phys. Lett. B} \textbf{\bibinfo{volume}{603}},
  \bibinfo{pages}{163} (\bibinfo{year}{2004}).


\bibitem[{\citenamefont{Nakatsukasa
  et~al.}(1999{\natexlab{a}})\citenamefont{Nakatsukasa, Walet, and
  Dang}}]{article:NakatsukasaWaletDang1999}
\bibinfo{author}{\bibfnamefont{T.}~\bibnamefont{Nakatsukasa}},
  \bibinfo{author}{\bibfnamefont{N.~R.} \bibnamefont{Walet}}, \bibnamefont{and}
  \bibinfo{author}{\bibfnamefont{G.~D.} \bibnamefont{Dang}},
  \bibinfo{journal}{Phys. Rev. C} \textbf{\bibinfo{volume}{208}},
  \bibinfo{pages}{90} (\bibinfo{year}{1999}{\natexlab{a}}).

\bibitem[{\citenamefont{Almehed and Walet}(2002)}]{article:AlmehedWalet2002}
\bibinfo{author}{\bibfnamefont{D.}~\bibnamefont{Almehed}} \bibnamefont{and}
  \bibinfo{author}{\bibfnamefont{N.~R.} \bibnamefont{Walet}},
  \bibinfo{journal}{Acta Physica Polonica B} \textbf{\bibinfo{volume}{34}},
  \bibinfo{pages}{2227} (\bibinfo{year}{2002}).

\bibitem[{\citenamefont{Kobayashi
  et~al.}(2005)\citenamefont{Kobayashi, Nakatsukasa, Matsuo, and 
  Matsuyanagi}}]{article:KNMM2005}
\bibinfo{author}{\bibfnamefont{M.} \bibfnamefont{Kobayasi}}, 
\bibinfo{author}{\bibfnamefont{T.} \bibfnamefont{Nakatsukasa}}, 
\bibinfo{author}{\bibfnamefont{M}. \bibfnamefont{Matsuo}}, 
\bibinfo{author}{\bibfnamefont{K}. \bibfnamefont{Matsuyanagi}},
\bibinfo{journal}{Prog.Theor.Phys}. \textbf{\bibinfo{volume}{113}},
\bibinfo{pages}{129-152} (\bibinfo{year}{2005}).



\bibitem[{\citenamefont{Nakatsukasa
  et~al.}(1999{\natexlab{b}})\citenamefont{Nakatsukasa, Walet, and
  Dang}}]{letter:NakatsukasaWaletDang1999}
\bibinfo{author}{\bibfnamefont{T.}~\bibnamefont{Nakatsukasa}},
  \bibinfo{author}{\bibfnamefont{N.~R.} \bibnamefont{Walet}}, \bibnamefont{and}
  \bibinfo{author}{\bibfnamefont{G.~D.} \bibnamefont{Dang}},
  \bibinfo{journal}{J. Phys. G} \textbf{\bibinfo{volume}{21}},
  \bibinfo{pages}{23} (\bibinfo{year}{1999}{\natexlab{b}}).

\bibitem[{\citenamefont{Klein et~al.}(1994)\citenamefont{Klein, Walet, and
  Dang}}]{article:KleinWaletDang1994}
\bibinfo{author}{\bibfnamefont{A.}~\bibnamefont{Klein}},
  \bibinfo{author}{\bibfnamefont{N.~R.} \bibnamefont{Walet}}, \bibnamefont{and}
  \bibinfo{author}{\bibfnamefont{G.~D.} \bibnamefont{Dang}},
  \bibinfo{journal}{Phys. Rev. B} \textbf{\bibinfo{volume}{322}},
  \bibinfo{pages}{11} (\bibinfo{year}{1994}).


\bibitem[{\citenamefont{Kobayasi et~al.}(2003)\citenamefont{Kobayasi,
  Nakatsukasa, Matsuo, and
  Matsuyanagi}}]{article:KobayasiNakatsukasaMatsuoMatsuyanagi2003}
\bibinfo{author}{\bibfnamefont{M.}~\bibnamefont{Kobayasi}},
  \bibinfo{author}{\bibfnamefont{T.}~\bibnamefont{Nakatsukasa}},
  \bibinfo{author}{\bibfnamefont{M.}~\bibnamefont{Matsuo}}, \bibnamefont{and}
  \bibinfo{author}{\bibfnamefont{K.}~\bibnamefont{Matsuyanagi}},
  \bibinfo{journal}{arXiv:nucl-th} \textbf{\bibinfo{volume}{030451}},
  \bibinfo{pages}{\,} (\bibinfo{year}{2003}).


\bibitem[{\citenamefont{Mang et~al.}(1976)\citenamefont{Mang, Samadi, and
  Ring}}]{article:MangSamadiRing1976}
\bibinfo{author}{\bibfnamefont{H.}~\bibnamefont{Mang}},
  \bibinfo{author}{\bibfnamefont{B.}~\bibnamefont{Samadi}}, \bibnamefont{and}
  \bibinfo{author}{\bibfnamefont{P.}~\bibnamefont{Ring}}, \bibinfo{journal}{Z.
  Phys.} \textbf{\bibinfo{volume}{A279}}, \bibinfo{pages}{325}
  (\bibinfo{year}{1976}).

\bibitem[{\citenamefont{Hill and Wheeler}(1953)}]{article:HillWheeler1953}
\bibinfo{author}{\bibfnamefont{D.~L.} \bibnamefont{Hill}} \bibnamefont{and}
  \bibinfo{author}{\bibfnamefont{J.~A.} \bibnamefont{Wheeler}},
  \bibinfo{journal}{Phys. Rev.} \textbf{\bibinfo{volume}{89}},
  \bibinfo{pages}{1102} (\bibinfo{year}{1953}).

\end{thebibliography}
%%%%%% % Produces the bibliography via BibTeX.

\end{document}